\documentclass[11pt]{article}

\usepackage{tcolorbox}
\tcbuselibrary{listings,breakable}
\usepackage[dvipsnames]{xcolor}
\newenvironment{promptbox}[1]{%
  \begin{tcolorbox}[
    colback=yellow!10!white,
    colframe=gray!75!black,
    title=#1,
    breakable,
    sharp corners,
    boxrule=0.6pt,
    left=6pt,right=6pt,top=6pt,bottom=6pt
  ]%
}{\end{tcolorbox}}
\usepackage{tikz}
\usepackage{pgfplots}
\usepgfplotslibrary{polar}
\pgfplotsset{compat=1.18}

\usepackage{amssymb}

\usepackage{makecell}

\usepackage[final]{acl}
\usepackage{float}

\usepackage{microtype}

\usepackage[english,bidi=default]{babel} 
\babelfont{rm}{TeXGyreTermesX} 

\usepackage{graphicx}
\usepackage{pifont}
\title{\textit{Graph of Trace}: Visualizing Execution Traces of Scientific Agents}

\author{
\textbf{Tianci Gao}$^{1,3}$%
\thanks{Equal contribution. lihaoxuan@sqz.ac.cn}
\quad
\textbf{Haoxuan Li}$^{1,2}$%
\footnotemark[1]
\quad
\textbf{Jianhe Li}$^{4}$
\quad
\textbf{Tianxiang Zhao}$^{1,5}$ \\
\textbf{Runze Shi}$^{2,1}$
\quad
\textbf{Weiran Wang}$^{6}$
\quad
\textbf{Zezhao Wu}$^{2}$
\quad
\textbf{Lu Mi}$^{2,1}$%
\thanks{Corresponding author. milu@mail.tsinghua.edu.cn} \\
$^{1}$Shanghai Qi Zhi Institute \quad
$^{2}$Tsinghua University \quad
$^{3}$Renmin University of China \\
$^{4}$Beihang University \quad
$^{5}$Georgia Institute of Technology \quad
$^{6}$Fudan University
}

\begin{document}

\maketitle
\begin{abstract}

Scientific AI agents can autonomously carry out complex research workflows, yet these unfolded workflows often remain difficult for humans to inspect and review, limiting interpretable, controllable and effective human–AI collaboration. 
To address this challenge, we present a monitoring and visualization framework that records fine-grained execution events and organizes them into a directed graph that makes agent workflows explicit as they proceed\footnote{We present our online demo, demo video and open sourced code at \href{https://github.com/NeuroAIHub/Graph-of-Trace-Visualizing-Execution-Trace-of-Scientific-Agents}{https://github.com/NeuroAIHub/Graph-of-Trace-Visualizing-Execution-Trace-of-Scientific-Agents}.}. 
The system records intermediate steps (e.g. tool calls and code executions), and renders them as real-time updated visual traces that expose workflow structure. 
This allows \textcolor{black}{users} to examine how results are produced, identify where failures emerge, and better understand agent behavior across different stages of the research process.
We conduct an evaluation on complex research tasks with domain experts of interdisciplinary backgrounds in AI, neuroscience, and biology. Experts report that structured traces visualization improves understanding of agent workflows, perceived interpretability, and usability for analysis and further interaction.

\end{abstract}
\section{Introduction}
Recent advances in large language models, when coupled with tool use and executable environments, have enabled autonomous agents capable of performing complex scientific research tasks ~\cite{ren_towards_2026, ding_scitoolagent_2025}. 
Scientific research is a long-horizon, multi-stage, and highly compositional process comprising multiple interdependent subtasks, such as literature review~\cite{huang2025deep}, experimental design~\cite{ren_towards_2026}, code implementation~\cite{novikov_alphaevolve_2025}, tool execution~\cite{ding_scitoolagent_2025}, data analysis~\cite{chen2024scienceagentbench}, and result interpretation~\cite{schmidgall2025agent}.

\textcolor{black}{
However, most existing works pursue an end-to-end framework, aiming at generating the final results or reports without human intervention \cite{novikov_alphaevolve_2025}. 
Errors introduced in early stages can propagate across dependent components, making late-stage correction costly or impractical \cite{sun_scienceboard_2025}. 
Thus, scientists' involvement also remains essential in these scientific workflows~\cite{Bellos_2025_ICCV}: refining research goals, validating assumptions, and ensuring alignment with research quality.
To support such involvement, scientists need to maintain a clear understanding of the workflow, review intermediate decisions, and intervene when necessary, rather than only receiving the final report at the end.
}

\begin{figure*}[t]
  \includegraphics[width=\textwidth]{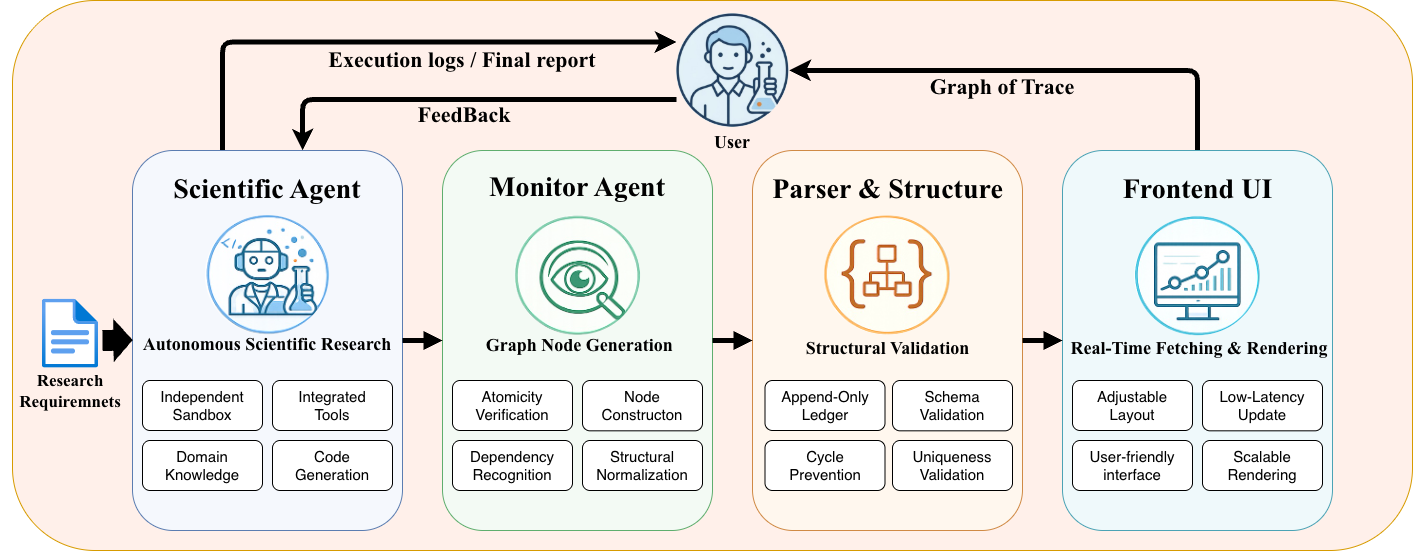}
  \caption{The overview of the framework. The architecture consists of four separate modules: the Scientific Agent performs autonomous task execution; the Monitor Agent receives operations and generates nodes for the garph; the Parser \& Structure enforces structural validation; then the Frontend UI renders the \textit{Graph of Trace} for real-time visualization.}
  \label{fig:overview}
\end{figure*}

To fill this gap, we introduce a \textcolor{black}{modularized} monitoring framework that records fine-grained execution events and structures them into an explicit directed graph of the agent's workflow, shown in Fig~\ref{fig:overview}. First, the Scientific Agent carries out research tasks and produces results, and a separate Monitor Agent reconstructs these outputs into normalized structural traces with explicit dependency relations. The resulting trace candidates are then passed to a parser and persistence module, which enforces schema-level constraints and appends only valid nodes and edges to construct \textit{\textit{Graph of Trace}}. On top of this, the visualization module continuously reflects the evolving graph, allowing users to inspect the research trajectory in real time without interfering with execution itself. By exposing execution traces rather than only outcomes, our approach supports process-level following, reviewing, and timely human \textcolor{black}{monitoring} throughout the research process, thereby advancing trustworthy AI-driven scientific research systems.
Our work introduces the following contributions:

1. \textbf{Conceptual Framework.} We introduce structured execution traces as an abstraction for representing agent workflows in an interpretable and analyzable form.

2. \textbf{System Design.} We develop a monitoring architecture that captures agent actions and constructs a real-time structured execution trace representation.

3. \textbf{UI Frontend Design.} We design an interactive interface that visualizes structured traces, enabling intuitive reviewing of execution processes.

\section{Related Works}

\subsection{Reasoning Representation and Visualization Frameworks}

Previous work has explored structured representations of LLM reasoning to improve interpretability beyond linear prompt traces. 
Graph-of-Thought \cite{Besta_2024} generalizes Chain-of-Thought~\cite{wei2022chain} and Tree-of-Thought~\cite{yao2023tree} paradigms by modeling reasoning as a directed graph of intermediate states, enabling non-linear branching, merging, and reuse of partial reasoning results. 
Building upon this paradigm, ReasonGraph \cite{li2025reasongraphvisualisationreasoningpaths} provides a web-based interface for visualizing and analyzing reasoning graphs generated by LLMs. DeepRare \cite{zhao_agentic_2026} traces LLM's reasoning process for healthcare tasks to improve the transparency and verifiability of the agent system.

These efforts primarily focus on revealing and inspecting the internal reasoning structure of LLM at the prompt level, instead of the intermediate behavioral actions of agentic systems.
Inspired by Graph-of-Thought style visualization, we propose a framework designed to represent and record the full execution process of autonomous scientific agents, instead of revealing the reasoning process.

\subsection{Autonomous Scientific Agents and Research Platforms}

Recent systems have substantially advanced the autonomy of scientific agents across the research lifecycle. 
Representative efforts include \textit{The AI Scientist} \cite{lu2024aiscientistfullyautomated}, \textit{Robin} \cite{ghareeb2025robinmultiagentautomatingscientific}, \textit{SciSciGPT} \cite{shao_sciscigpt_2025}, \textit{Biomni} \cite{huang_biomni_2025}, and \textit{OpenLens AI}\cite{cheng2025openlensaifullyautonomous}, among others. 
Collectively, this line of work establishes autonomous scientific discovery as a rapidly maturing paradigm， which demonstrates the increasing feasibility of long-horizon, tool-integrated, and domain-aware scientific automation, spanning hypothesis generation, experimentation, analysis, and reporting within end-to-end or multi-agent research workflows. 


However, most platforms are targeted at exposing final research results without structured, dependency-aware representations of intermediate steps, which limits the capabilities to follow execution trajectories, review intermediate states, and interact via real-time intervention, even though outcome-level evaluation is supported.

\section{Framework}

\subsection{System Overview}





For explicitly and correctly visualizing the research process conducted by agents, we design a four-module framework shown in Figure~\ref{fig:overview}, which consists of (1) a Scientific Agent for autonomous execution, (2) a Monitor Agent for reconstructing structural traces, (3) a parser and persistence module that enforces structural contracts, and (4) a visualization module that exposes the evolving \textit{Graph of Trace} in real time. 

\subsection{Scientific Agent}

The Scientific Agent is responsible for task execution and result production. In our implementation, we use OpenHands~\cite{wang2025openhandssoftwareagentsdk} as a representative autonomous agent framework. 
\textcolor{black}{
The agent operates autonomously and reports completed subtasks through MCP calls. 
}
\textcolor{black}{
It is important to note that the Scientific Agent does not construct or maintain the \textit{Graph of Trace} directly. Instead, the construction of \textit{Graph of Trace} is handled independently by other modules, ensuring that the framework itself is not tightly coupled to any agent platform.
}

\subsection{Monitor Agent}



The Monitor Agent integrates the information reported by the Scientific Agent into standardized \textit{Graph of Trace} nodes according to predefined node schemas. It further determines parent--child relationships among nodes, ensuring structural consistency during graph construction.
This separation between task execution and \textit{Graph of Trace} construction avoids introducing irrelevant or redundant information into the Scientific Agent’s working context, providing a cleaner and more reliable representation of the research process.
To enable plug-and-play adaptation to any scientific agents or workflows, the Monitor is encapsulated as an MCP-compliant tool \cite{Hou_Model_2026}, a protocol-level interface rather than integrating it into the execution framework.
The detailed prompt of the agent is attached in the Appendix~\ref{sec:appendix}.

\subsection{Parser and Structure}
We define the update process of \textit{Graph of Trace} as an append-only JSON ledger. 
This append-only design prevents retrospective modifications which could introduce context noises, thus preserving the authenticity of recorded subtasks.

Within this ledger, each node represents a single atomic subtask and contains structured metadata, including a textual description, references to artifacts, and identifiers of the parent nodes. Directed edges capture dependencies between nodes, representing causal or logical progression rather than mere temporal order.
\textcolor{black}{
Before integrating candidate nodes provided by the Monitor Agent, the Parser enforces schema-constrained validation: nodes must conform to the predefined structure, parent references must exist, and cycles or orphaned dependencies are not allowed. 
}
Only nodes and edges that pass validation are appended, maintaining structural integrity and the monotonic growth of the \textit{Graph of Trace}.

\begin{figure*}[h]
    \centering
    \includegraphics[width=\textwidth]{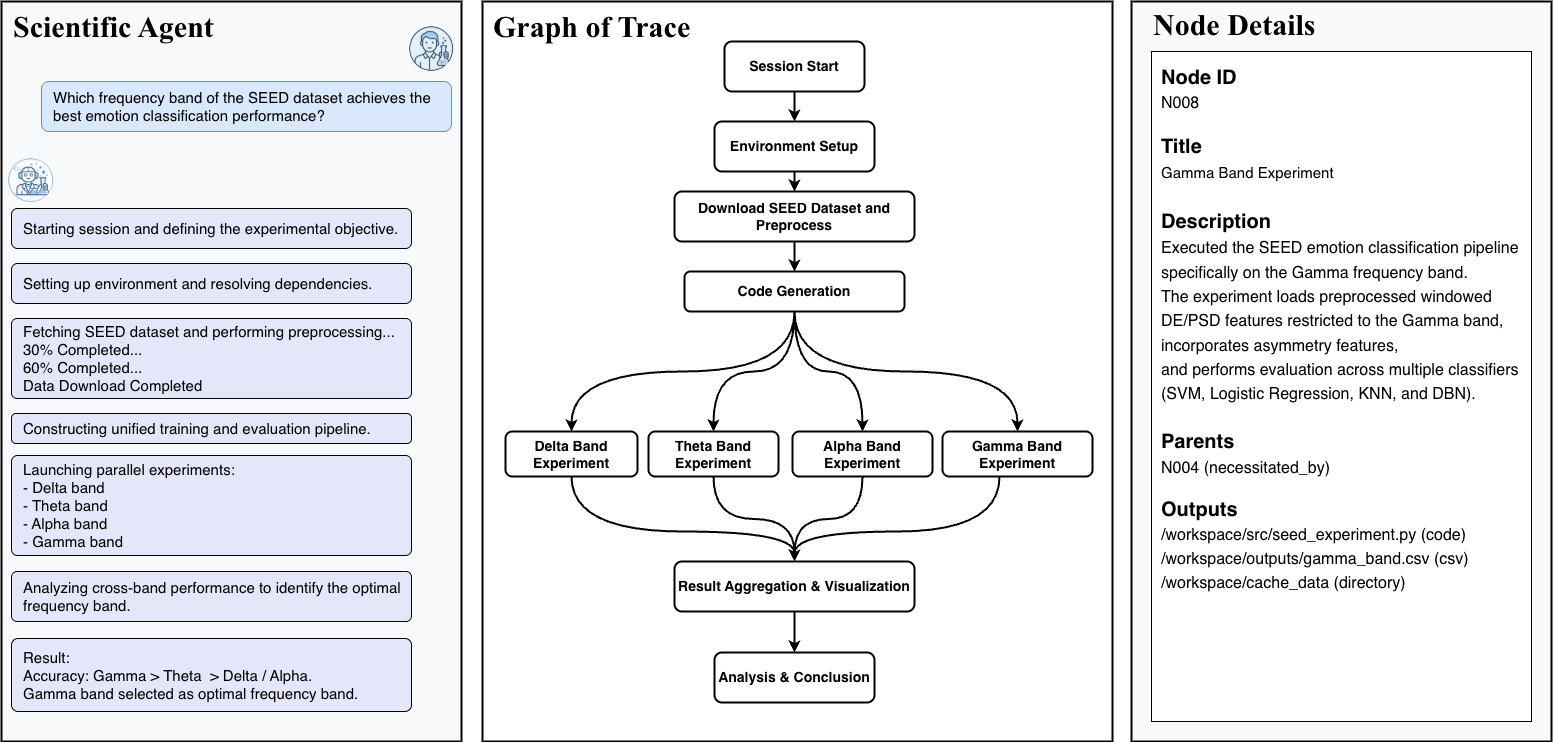}
    \caption{Illustration of UI Design: From left to right: (1) Scientific Agent Conversational Panel. Users interact with the scientific agent through natural language dialogue, while the agent’s stepwise responses reflect the progression of the task; (2) 
    \textit{Graph of Trace} Visualization. The agent’s workflow is rendered as a layered, top-down directed acyclic graph, beginning with session initialization, followed by environment setup, data preprocessing, model implementation, parallel experiments, and concluding analysis; (3) Node Details. Selecting a node reveals its structured metadata, including identifier, title, description, parent dependencies, and generated intermediate outputs.
    }
  \label{fig:ui}
\end{figure*}
\begin{figure*}[t!]
  \includegraphics[width=0.97\textwidth]{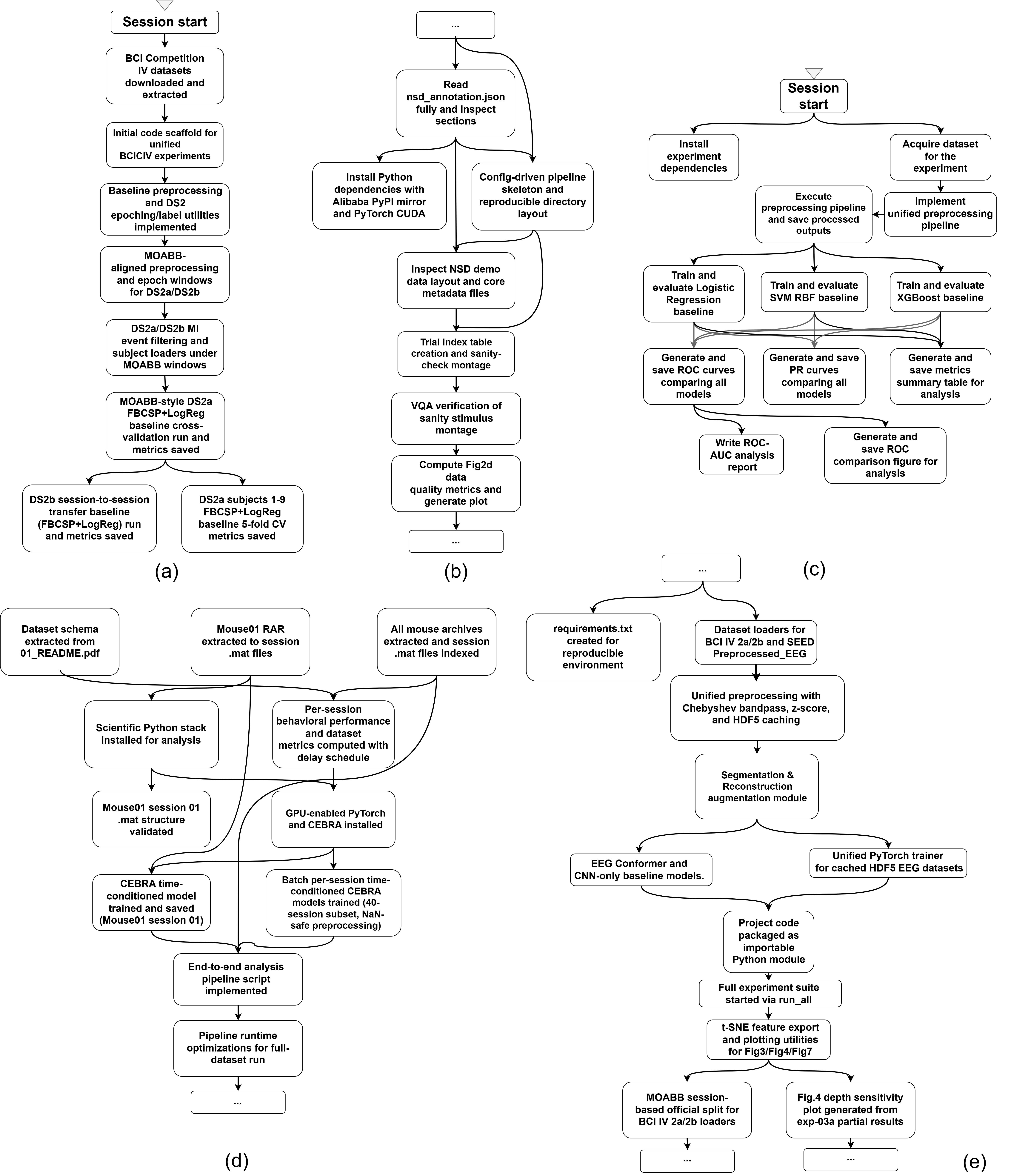}
  \caption{The demonstration of 5 example cases of \textit{Graph of Trace} in the scientific research studies, ranging from neuroscientific discovery to brain-computer interface applications. The selected tasks (a)\textasciitilde (e) include dataset preparation, Python runtime environment setup, data analysis, comparisons of machine learning methods, reproduction of a certain experiment, etc. These tasks vary in methodological complexity, thus resulting in the diversity of structures and complexity of \textit{Graph of Trace}. Note that the block containing an ellipsis (…) represents omitted parts of the execution graph due to space limit.}
  \label{fig:case}
\end{figure*}

\subsection{Frontend Visualization}
\textcolor{black}{
The Frontend Visualization module renders the \textit{Graph of Trace} as a directed graph in which nodes correspond to atomic subtasks and edges encode declared dependencies. Users can inspect node-level metadata, including descriptions, artifacts, and dependency relations, enabling localized auditing of specific execution steps.
}
The visualization module remains synchronized with the persistent \textit{Graph of Trace} state, ensuring consistency between backend structure and frontend rendering while preserving a clear separation between operational execution and interpretative transparency.

\section{UI Design}

The interface is organized into three coordinated panels to support interaction, monitoring, and inspection of the agent workflow, shown in Figure~\ref{fig:ui}.

The left panel integrates a conversational interface through which users interact directly with the Scientific Agent. In addition to the dialog history, this panel exposes the agent’s step-by-step execution process, including actions such as tool invocation, code generation, and command execution. This design allows users to follow the agent’s ongoing behavior in a familiar chat-based setting.

The center panel presents the \textit{Graph of Trace}, which is rendered with a React-based\footnote{\url{https://legacy.reactjs.org/}} frontend. This view visualizes the structural relations among the agent’s operations and provides an overview of the workflow as it evolves. Users can freely drag nodes, zoom in and out, and pan across the graph for flexible exploration. A minimap is also provided to support navigation in larger traces.

The right panel provides a detailed inspection for the selected node. When a node is clicked, the interface displays its full operation description, its connected parent nodes, and the corresponding intermediate output produced by the agent at that step. This enables users to examine local execution details while maintaining awareness of the broader workflow structure.


\section{Evaluation}
To evaluate the practical effectiveness of the proposed framework, five domain experts with interdisciplinary backgrounds in AI, neuroscience, and biology conducted a human-centered assessment focusing on real-world scientific replication. 
\textcolor{black}{
Each expert selected one or two representative research papers from their respective domains. They then reproduced the core experiments and workflows using the scientific agent platform integrated with our framework.
}
Overall, we conducted seven case studies of different scientific research that vary substantially in methodological complexity. The generated graphs have diverse structures and different levels of complexity, with the number of nodes ranging from as few as five to more than twenty. 

In Figure~\ref{fig:case}, we have shown \textit{Graph of Trace} in 5 different example scientific research cases. For case (a), the agent is asked to perform a structured exploration of brain computer interface (BCI) Competition IV Dataset under a standardized protocol \cite{tangermann_review_2012}. The \textit{Graph of Trace} is generally linear in sequence but exhibits parallel operations, resulting in a relatively simple structure. 
For case (b), it requires the agent to perform analysis like long-term recognition memory effects and representational similarity analysis on the Natural Scenes Dataset (NSD) \cite{allenmassive2022}. The \textit{Graph of Trace} is also roughly linear in sequence but exhibits more complex parallel and hierarchical operations and dependencies. 
\textcolor{black}{
In case (c), we asked the agent to conduct a simple comparison of machine learning methods which are commonly applied in data-driven scientific discovery. The structure of its \textit{Graph of Trace} diverges into several parallel model evaluation branches, including Logistic Regression, SVM with RBF kernel, and XGBoost. It further coordinates cross-model comparison procedures, generating curve analyses, summary metric tables and comparison figures.  
}
For case (d), the agent is required to investigate how reward-related representations in hippocampal neuronal activity evolve over extended experience, based on the predictive coding of reward in the Hippocampus dataset \cite{yaghoubi2026predictive}. The \textit{Graph of Trace} is more complicated, with more parallel operations and long-range planning. In case (e), we asked the agent to reproduce the baseline motor imagery (MI) experiments from BCI Competition IV \cite{tangermann_review_2012}. Its \textit{Graph of Trace} has a structure from which we can see a research process proceeding from the initial preparation, to a phase of exploration, then to focused execution, and finally branching into specific directions. 
\textcolor{black}{
Beyond the cases shown in the figure, in case (f), we asked the agent to analyze the SEED dataset and identify which frequency band contains more important discriminative information for emotion recognition \cite{Zheng2015SEED}. In the \textit{Graph of Trace}, we can observe four parallel branches, each of which trains and evaluates a classification model on a different EEG siganl frequency band (alpha, beta, gamma, and theta). The final node aggregates the results from all branches and concludes that the gamma and theta bands achieve the highest classification accuracy. 
For case (g), the agent conducted a exploration of the BCI Competition IV Dataset (DS1, DS2a, and DS2b) \cite{tangermann_review_2012},  including benchmark evaluation, cross-session transfer analysis, EOG artifact robustness testing, and synthetic data validation. 
Its \textit{Graph of Trace} follows a mainly linear structure, including dataset preparation, environment setup, preprocessing, multi-dataset experiments, artifact removal comparison, and result visualization. The overall \textit{Graph of Trace} clearly reconstructs the complete research workflow.
}

We further analyze some failure cases observed in practical use. 
When the Scientific Agent completes multiple subtasks without immediately reporting them to the Monitor Agent through MCP calls, some operations are not reflected in the \textit{Graph of Trace} in real time. However, the Scientific Agent usually reports the left work in the next MCP call. Therefore, this problem mainly appears as delayed updates rather than permanent information loss.

The second issue is incomplete dependency modeling by the Monitor Agent. In some cases, prerequisite steps such as environment setup or literature review are not correctly identified as parent nodes of downstream experiments. Similarly, during final result summarization, multiple experimental branches may not be fully connected to the final report node, such as task (c) in \ref{fig:case}. 

The third issue appears when the Scientific Agent starts exploring a new branch after long execution. Because of the long context, the Monitor Agent may fail to identify the correct starting parent node and directly connects the new branch to the last node of the previous branch. This creates incorrect dependencies between branches and makes parallel or independent workflows appear sequential.


\begin{table}[h]
\centering
\small
\begin{tabular}{>{\centering\arraybackslash}p{0.05\linewidth}cc>{\raggedright\arraybackslash}p{0.03\linewidth}>{\raggedright\arraybackslash}p{0.03\linewidth}>{\raggedright\arraybackslash}p{0.07\linewidth}l|l}
\hline
Case& \makecell{Node \\ (Acc)} & \makecell{Edge \\ (Acc) }
& Lat.& Int. 
& Aesth. 
& Usab.  & Baseline\\
\hline

1 & 23/23 & 30/32 & 4 & 2 & 4 & 2 & 0.8\\
2 & 9/9 & 8/8 & 5 & 5 & 5 & 5 & 2.0\\
3 & 21/21 & 26/26 & 5 & 5 & 3 & 4 & 1.6\\
4 & 18/18 & 19/21 & 5 & 4 & 4 & 5 & 4.0\\
5 & 5/6 & 4/5 & 4 & 5 & 5 & 4 & 3.2\\
6 & 12/13 & 15/15 & 5 & 5 & 4 & 5 & 3.0\\
7 & 13/13 & 18/18 & 5 & 5 & 4 & 5 & 3.0\\

\hline
Avg. 
& 0.98 & 0.96 
& 1.6s & 4.43 
& 4.14 
& 4.29 & 2.51\\

\hline
\end{tabular}
\caption{Summary of expert evaluation results across seven experimental cases. 
\textcolor{black}{
Node/Edge (Acc): accurate over total; 
Lat.: We compute the average latency by taking the midpoint of the selected delay interval and averaging these midpoint values across all cases;
Int.: Interpretability; 
Aesth.: Aesthetics; Usab.: Usability; 
Baseline: Usability w/o \textit{Graph of Trace}.
}
}
\label{Evaluation Score}
\end{table}

Across all cases, we evaluated the quality of the visualization with the following aspects, 

\noindent\textcolor{black}{
\textbf{Structural Validity.} Throughout the evaluation, we observed no formatting or graph rendering errors, indicating that the visualization faithfully preserves the underlying execution structure.
}

\noindent\textbf{Node-Level Fidelity.} The overall node correctness ratio (correct nodes / total nodes) was 0.98, with a mean per-expert accuracy of 0.96 (averaged over individual experts).

\noindent\textbf{Relation-Level Fidelity.} The overall edge correctness ratio (correct edges / total edges) was 0.96, with a mean per-expert accuracy of 0.94 (averaged over individual experts).

\noindent\textbf{Temporal Consistency.} The system also demonstrated strong consistency with graph updates tightly synchronized to agent execution and a low average latency of 1.6 seconds.

We also collected the users' feedback with the following ratings on a five-point Likert scale (1 = lowest, 5 = highest). 

\noindent\textbf{Visual Aesthetics.} The interface received an average score of 4.14, noting that the graph remained clear and readable throughout the whole process. 

\noindent\textbf{Expert Interpretability.} The system achieved a mean rating of 4.43, reflecting that the visualization effectively translated complex agent reasoning into a structured representation that experts could easily follow and review. 

\noindent\textbf{Usability.} The average score reached 4.29. In contrast, when experts were asked to understand the agent’s behavior solely by reading raw dialogues, execution logs, and source code, the average usability rating dropped to 2.51. This comparison indicates that the structured visualization substantially reduced cognitive burden and improved the efficiency of reviewing.

Overall, the evaluation results indicate that in stress tests involving long-horizon, multi-step scientific workflows, \textit{Graph of Trace} enabled real-time auditing of agent behavior. 
Therefore, the framework effectively bridges raw execution logs and human-interpretable scientific reasoning. 

\section{Applications}

The proposed framework establishes a practical basis for supervising autonomous agent behavior in complex scientific settings and lays the groundwork for more convenient human--agent collaboration. First, the system provides real-time overview for research workflows. By transforming long-horizon sequential execution records into a structured \textit{Graph of Trace}, the framework enables experts to inspect multi-stage experimental process more clearly rather than just a final report. Furthermore, the explicit presentation of the structured research workflow and its corresponding outputs facilitates the verification and reproducibility of research processes.

\section{Conclusion}
In this work, we presented a monitoring and visualization framework that establishes a collaborative tracing protocol between the Scientific Agent and a dedicated Monitor. 
The Monitor Agent then synthesizes these reports into coherent \textit{Graph of Trace} nodes, constructing a real-time, interpretable representation of the research workflow.

Our human-centered evaluation demonstrates that this structured trace visualization significantly enhances expert understanding, perceived interpretability, and usability compared to traditional end-to-end outputs. 
Ultimately, this framework bridges the gap between autonomous capability and scientific reliability. By facilitating active reporting and structured visualization, we pave the way for more trustworthy, controllable, and collaborative human-AI scientific discovery systems.

In future work, we plan to extend the system with node-level backtracking, enabling experts to revisit earlier states in the workflow and give more precise and in-time intervention while monitoring the agent trajectory in real time, thereby preventing error accumulation across subsequent steps. We also plan to conduct a systematic evaluation of different scientific agents by using structured visualizations to compare how they carry out complex scientific research tasks. It will further support the development of domain-specific benchmarks grounded in observable research trajectories. Finally, we plan to study the interaction between human expert feedback on the \textit{Graph of Trace} and the trajectories produced by agents, with the goal of identifying new patterns of human–AI collaboration in scientific research.

\section{Limitations}
Despite the promising results, several limitations remain that warrant future investigation.

\textbf{Reliance on Agent Compliance.} 
The framework assumes that the Scientific Agent consistently invokes the MCP tracing tools at appropriate semantic boundaries. If tracing calls are skipped or delayed, the resulting Graph of Trace can become incomplete or fragmented. As a result, trace quality currently depends on the agent’s compliance with the instrumentation protocol. 

\textbf{Monitor Interpretation Errors.} 
The framework uses an LLM-based interpretation module to convert low-level execution reports into higher-level subtask units. Although we apply structured prompts and output validation to improve reliability, this module can still misinterpret ambiguous reports or introduce hallucinated structure. Such errors may lead to incorrect task decomposition or inaccurate dependency links.

\textcolor{black}{
\textbf{Scalability of Large Execution Graphs.}
For extremely long workflows that may generate hundreds of nodes, the current framework still lacks richer interaction mechanisms such as hierarchical subgraph collapsing, semantic module grouping, and progressive node expansion. This can increase the cognitive burden on users when inspecting long-term tasks.
}

\bibliography{custom}

\clearpage

\appendix
\section{Appendix}
\label{sec:appendix}
\subsection{Prompts}

The following prompt is designed for the Monitor Agent, which is to receive execution information from the Scientific Agent and generate structured execution nodes. To ensure generation correctness, the prompt explicitly defines the schema and semantics of a node.
In addition to node generation, a critical responsibility of the Monitor is to establish parent-child relationships between nodes. The Monitor should choose a valid, logically consistent, and existing node as the parent. We therefore define explicit rules in the prompt to constrain this process.
To further improve reliability, we include representative examples in the prompt. In our experiments, these examples significantly improve the Monitor’s performance.

\begin{promptbox}{Prompts: Monitor}
\ttfamily
You are a strict information extractor. Convert the provided subtask summary into one or more DAG nodes.\\

<Node Definition>\\
A node represents a research-relevant operation that contributes to the experimental, analytical, or infrastructural progression of the research.\\

<Node Boundary and Splitting Rules>\\
Split actions into separate nodes IF AND ONLY IF:\\
1. One action explicitly depends on the result of another.\\
2. Each action has independent research-level semantic significance.\\
...

<Parent Selection Rules>\\
- Each node MUST have at least one parent.\\
- Parents represent logical research justification, NOT chronological execution order.\\
- You MUST select parent ids from existing nodes.\\
...

<Artifacts Requirements>\\
Each node MUST include an artifacts list (may be empty only if no concrete file artifact exists).\\
Artifacts represent the verifiable output of the node.\\
Rules:\\
- Execution, visualization, analysis, and conclusion nodes MUST produce concrete, inspectable artifacts.\\
- Setup or literature nodes MAY have empty artifacts if no file-level output exists.\\

<Examples>\\
...\\
<Previous \textit{Graph of Trace}>\\
...\\
<Output Schema>\\
...\\
\end{promptbox}


We provide an MCP description to guide the Scientific Agent on how to call the Monitor Agent.
To maximize information flow into the Monitor, we encourage the Scientific Agent to invoke the MCP after completing each subtask, ensuring that the Monitor receives better execution information, while avoiding unnecessary verbosity such as raw logs or debugging details.
Additionally, we also include several representative examples for timely and properly formatted Monitor Agent calls.

\vspace{1em}
\begin{promptbox}{Prompts: MCP Description}
\ttfamily
This tool records **what was done** (an executable, verifiable operation) and its artifact dependencies.\\
It is NOT for chain-of-thought, progress logs, explanations, or exploratory work.\\

<Call Gate>\\
* Each MCP call record MUST represent exactly ONE executable action step.\\
* Do NOT call for planning, debugging, or rephrasing.\\
* After completing any subtask, you must immediately record it via the MCP tool.\\
* When in doubt about whether the step qualifies or not, you must still record it.\\
...

<Input Requirements>\\
* Provide exactly ONE subtask describing a minimal executable unit with a factual `title` and `description`.\\
* If this subtask depends on prior results/steps, provide a dependence list to describe those dependencies.\\
* Dependence list MAY include multiple items.\\
...\\

<EXAMPLES>\\
...
\end{promptbox}

\vspace{1em}

\vspace{1em}


\section{Expert Evaluation Questionnaire}
We design a comprehensive questionnaire and conduct a structured expert evaluation of the visualization framework, covering structural validity, node and edge level fidelity, latency, and usability. The results are reported in Table~\ref{Evaluation Score}.

\subsection{Structural Validity}

The times of failed rendering during execution: \_\_\_

\subsection{Node-Level Fidelity}

Total nodes: \_\_\_ \quad
Accurate nodes: \_\_\_ \quad

\subsection{Relation-Level Fidelity}

Total edges: \_\_\_ \quad
Correct edges: \_\_\_ \quad

\subsection{Temporal Consistency}

Update latency: \\
$\square$ Real-time ($<$2s) \quad 
$\square$ Slight (2--5s) \quad \\
$\square$ Moderate (5--10s) \quad 
$\square$ Severe (10--20s) \quad \\
$\square$ Unacceptable ($>$20s / unstable)

\subsection{Expert Interpretability}

$\square$ Immediate Understanding \quad \\
$\square$ Minor clarification needed \quad \\
$\square$ Partially with effort needed \quad \\
$\square$ Fragmented \quad \\
$\square$ No effective understanding

\subsection{Visual Aesthetics}

$\square$ Excellent \quad 
$\square$ Good \quad 
$\square$ Fair \quad
$\square$ Poor \quad \\
$\square$ Very poor

\subsection{Usability}

$\square$ Highly convenient \quad \\
$\square$ Convenient \quad \\
$\square$ Moderately convenient \quad \\
$\square$ Inconvenient \quad \\
$\square$ Very inconvenient \\

\end{document}